\documentclass[preprintnumbers,amsmath,amssymb,floatfix,11pt,prd,onecolumn,
superscriptaddress,nofootinbib]{revtex4}
\usepackage{latexsym}
\usepackage{epsfig}
\usepackage{amsmath}
\usepackage{epstopdf}
\usepackage{amssymb}
\usepackage{graphicx}
\begin{document}

\title{\bf Dynamics of a Charged Particle Around a Slowly Rotating Kerr
 Black Hole Immersed in Magnetic Field }

\author{ Saqib Hussain }
\email{s.hussain2907@gmail.com} \affiliation{School of Natural
Sciences (SNS), National University of Science and Technology
(NUST), H-12, Islamabad, Pakistan}

\author{Ibrar Hussain}
\email{ibrar.hussain@seecs.nust.edu.pk} \affiliation{School of
Electrical
 Engineering and Computer Science (SEECS),
 National University of Sciences and Technology (NUST), H-12, Islamabad, Pakistan}

\author{Mubasher Jamil}
\email{mjamil@sns.nust.edu.pk ,
jamil.camp@gmail.com}\affiliation{School of Natural Sciences (SNS),
National University of Science and Technology (NUST), H-12,
Islamabad, Pakistan}

\begin{abstract}
{\bf Abstract:} The dynamics of a charged particle moving around a
slowly rotating Kerr black hole in the presence of an external
magnetic field is investigated. We are interested to explore the
conditions under which the charged particle can escape from the
gravitational field of the black hole after colliding with another
particle. The escape velocity of the charged particle in the
innermost stable circular orbit is calculated. The effective
potential and escape velocity of the charged particle with angular
momentum in the presence of magnetic field is analyzed. This work
serves  as an extension of a preceding paper dealing with the
Schwarzschild black hole [Zahrani {\it et al}, Phys. Rev. D 87,
084043 (2013)].
\end{abstract}
 \maketitle

\newpage
\section{Introduction}

The dynamics of particles (massive or massless, charged or neutral)
around a black hole is among the most important and interesting
problems of black hole astrophysics. These studies not only help us
to understand the geometrical structure of spacetimes but also shed
light on the high energy phenomenon occurring near the black hole
such as formation of jets (which involve particles to escape) and
accretion disks (particles orbiting in circular orbits). Due to the
presence of strong gravitational and electromagnetic fields, charged
particles in general do not follow stable orbits and inter-particle
collisions are most common. The aftermath of these collisions among
numerous particles lead to various interesting astrophysical
phenomenon.

There are numerous astrophysical evidence that magnetic field might
be present in the nearby surrounding of black holes \cite{new,black}
which support the large scale jets. These jets are most likely the
source of cosmic rays and high energy particles coming from nearby
galaxies. The origin of this magnetic field is the probable
existence of plasma in the vicinity of a black hole in the form of
an accretion disk or a charged gas cloud \cite{1,2}. The
relativistic motion of particles in the conducting matter in the
accretion disk can generate the regular magnetic field inside the
disk. Therefore near the event horizon of a  black hole, it is
expected that there exists much strong magnetic field. To an
approximation, it is presumed that this field does not effect the
geometry of the black hole but it does effect the motion of the
charged particles moving around the black hole \cite{p9,3}.

More importantly, a rotating black hole may provide sufficient
energy to the particle moving around it due to which the particle
may escape to spatial infinity. This physical effect appears to play
a crucial role in the ejection of high energy particles from
accretion disks around black holes. In the process of ejection of
high energy particles, besides the rotation of black hole, the
magnetic field plays an important role \cite{4,5}. Note that if the
black hole is carrying electric charge producing the static electric
field (also called Coulomb field), then the mere rotation of black
hole itself induces the magnetic field. Acceleration of the particle
by the black hole is generally explained in \cite{p6}. Other
interesting processes around black holes may include evaporation and
phantom energy accretion onto black holes \cite{jamil}.

During the motion of a charged particle around a magnetized black
hole, it remains under the influence of both gravitational and
electromagnetic forces which makes the situation complicated
\cite{p1,13}.  In the present article, it is considered that a
charged particle is orbiting in the innermost stable circular orbit
(ISCO) of a slowly rotating Kerr black hole and is suddenly hit by a
radially incoming neutral particle. The aftermath of collision will
depend on the energy of the incoming particle which may result one
of the three possible outcomes: charged particle may escape to
infinity; being captured by the black hole or keep orbiting in ISCO.
However predicting the nature of outcome is compounded by the facts
that particle is charged and interacts with the magnetic field and
is frame dragged by the Kerr black hole. It should be noted that the
present work is altogether different from the BSW mechanism where
two particles (with non-zero angular momentum and high energies)
arrive from spatial infinity and collide near the event horizon to
generate surplus energy in the center of mass frame \cite{bsw}. In
literature, motion of charged particles in ISCO around various black
holes has been studied (\cite{ref} and see therein).

Here we consider a slowly rotating Kerr black hole which is
surrounded by an axially symmetric magnetic field homogeneous at
infinity. Almost similar  problem was studied for weakly charged
rotating black holes in \cite{6}. Their main conclusion is that, if
the magnetic field is present than the ISCO is located closer to the
black hole horizon. In general, the effect of the black hole
rotation on the motion of a neutral particle  is same as the effect
of magnetic field on the motion of a charged particle \cite{11,p2}.

To study the escape velocity of a particle from the vicinity of
black hole, in this paper we first consider a neutral particle
moving around a slowly rotating Kerr black hole in the absence of
magnetic field and collides with another particle. For simplicity we
consider the motion in the equatorial plane only. Then we consider
the same problem for a charged particle in the presence of magnetic
field. We focus under what circumstances the particle can escape
from the strong gravitational field to infinity. Magnetic field is
homogeneous far from the black hole and gravitational field is
ignorable. Thus, far from the black hole charged particle moves in a
homogeneous magnetic field. If the magnetic field is absent then the
equations of motion are simple a little and can be solved
analytically. When a particle moving in a non uniform magnetic field
in the absence of black hole its motion is chaotic \cite{8,9}. We
are extending a previous work  \cite{11}  for the slowly rotating
Kerr black hole.

 The outline of the paper is as follows: In section II we
explain our model and derive an expression for escape velocity of
the neutral particle. In section III we derive the equations of
motion of the charged particle moving around a slowly rotating
weakly magnetized Kerr black hole. In section IV we give the
dimensionless form of the equations. Trajectories for escape energy
and escape velocity of the particle are discussed in section V and
VI respectively and their graphs are given in the appendix. Summery
and conclusion are presented in section VII. Throughout we use sign
convention $(+,-,-,-)$ and units where $c=1,G=1$.

\section{Escape Velocity For a Neutral Particle}

We start with the simple case of calculating the escape velocity
when the particle is neutral and magnetic field is absent. The Kerr
metric is given by \cite{15}
\begin{eqnarray}\label{1}
ds^{2}&=&\frac{\Delta-a^{2}\sin^{2}\theta}{\rho^{2}}dt^{2}+\frac{4Mar\sin^{2}\theta}{\rho^{2}}d\phi
dt-\frac{\rho^{2}}{\Delta}dr^{2}
-\rho^{2}d\theta^{2}-\frac{A\sin^{2}\theta}{\rho^{2}}d\phi^{2},
\nonumber\\&& \Delta\equiv r^{2}-2Mr+a^{2}, \ \ \rho^{2}\equiv
r^{2}+a^{2}\cos^{2}\theta, \ \
A\equiv(r^{2}+a^{2})^{2}-a^{2}\Delta\sin^{2}\theta,
\end{eqnarray}
where $M$ is the mass  and $a$ is the spin of the black hole and
interpreted as the angular momentum per unit mass of the black hole
$a=\frac{L}{M}$. The horizons of Kerr metric are obtained by solving
\begin{equation}
  \Delta(r)=r^{2}+a^{2}-2Mr=0.
\end{equation}
From the above equation we get two values of $r$:
\begin{equation}
  r_{+}=M+\sqrt{M^{2}-a^{2}}, ~~~r_{-}=M-\sqrt{M^{2}-a^{2}}.
\end{equation}
Note that $\Delta>0$ for $r>r_{+}$ and $r<r_{-}$  and $\Delta<0$ for
$r_{-}<r<r_{+}$ \cite{14}.
 The region $r=r_{+}$ represents the event horizon while $r_-$ is termed as the Cauchy horizon.
Further $r=0$ and $\theta=\frac{\pi}{2}$ is the location of a
curvature ring-like singularity in the Kerr spacetime.

In literature, slowly rotating Kerr black holes have been
investigated for numerous astrophysical processes including as
retro-MACHOS \cite{qadir}, particle acceleration via BSW mechanism
\cite{bsw1}, thin accretion disk and accretion rates \cite{lobo}, to
list a few. Hence we consider the slowly rotating black hole and
neglect the terms involving $a^{2}$. The line element in (\ref{1})
becomes
\begin{equation}\label{111}
ds^2=(1-\frac{r_{g}}{r})dt^2+\frac{4aM\sin^{2}\theta}{r}d\phi
dt-\frac{1}{1-\frac{r_{g}}{r}}dr^2-r^2d\theta^2-r^2\sin^2\theta
d\phi^2.
\end{equation}
Here $r_{g}=2M$, is the gravitational radius of the slowly rotating
Kerr black hole just like Schwarzschild black hole (Note that for a
slowly rotating Kerr and Schwarzschild black hole the horizon occurs
at $r=r_{g}$ ). Clearly the metric (\ref{111}) is stationary but
non-static since $dt\rightarrow-dt$, changes the signature of
metric. The metric is also axially symmetric (invariance under
$d\theta\rightarrow-d\theta$).

In terms of Lagrangian mechanics ($\mathcal{L}=g_{\mu\nu}\dot
x^\mu\dot x^\nu$), the $t$ and $\phi$ coordinates are cyclic which
lead to two conserved quantities namely energy and angular momentum
with the corresponding Noether symmetry generators
\begin{equation}\label{1a}
\xi_{(t)}= \xi_{(t)}^\mu\partial_\mu =\frac{\partial}{\partial t},
\qquad
\xi_{(\phi)}=\xi_{(\phi)}^\mu\partial_\mu=\frac{\partial}{\partial\phi}.
\end{equation}
This shows that the black hole metric is invariant under time
translation and rotation around symmetry axis. The corresponding
conserved quantities are the energy $\mathcal{E}$ per unit mass and azimuthal
angular momentum $L_{z}$ per unit mass \footnote{Given a Lagrangian
$\mathcal{L}=g_{\mu\nu}\dot x^\mu\dot x^\nu$, one can calculate the
conserved quantities corresponding to cyclic coordinates $t$ and
$\phi$ as $\frac{d}{d\tau}\frac{\partial \mathcal{L}}{\partial \dot
t}=0,$ and $\frac{d}{d\tau}\frac{\partial \mathcal{L}}{\partial
\dot\phi}=0,$ yielding $\frac{\partial \mathcal{L}}{\partial \dot
t}=\mathcal{E}\equiv -p_\mu \xi_{(t)}^\mu/m $, and $\frac{\partial
\mathcal{L}}{\partial \dot \phi}=L_z\equiv p_\mu
\xi_{(\phi)}^\mu/m$. Solving these equations simultaneously, one can
obtain (\ref{3}).}

\begin{align}\label{3}
  \dot{t}&=\frac{r^{3}\mathcal{E}+aL_{z}r_{g}}{r^{2}(r-r_{g})},\nonumber\\
  \dot{\phi}&=\frac{1}{r^{2}}\bigg(\frac{ar_{g}\mathcal{E}}{(r-r_{g})}+\frac{L_{z}}{\sin^{2}\theta}\bigg).
\end{align}

From the astrophysical perspective, it is known that particles orbit
a rotating black hole in the equatorial plane \cite{sand}. Therefore
we choose $\theta=\frac{\pi}{2}$ to get
\begin{align}\label{1.5d}
  \dot{t}&=\frac{r^{3}\mathcal{E}+aL_{z}r_{g}}{r^{2}(r-r_{g})},\nonumber\\
  \dot{\phi}&=\frac{1}{r^{2}}\big(\frac{ar_{g}\mathcal{E}}{(r-r_{g})}+L_{z}\big).
\end{align}
 Throughout in this paper the
over dot represents differentiation with respect to proper time
$\tau$.

Using the normalization condition, $u^{\mu}u_{\mu}=1$, we get the
equation of motion
\begin{equation}\label{4}
\dot{r}^{2}=\frac{(\mathcal{E} r^{2}-
aL_{z})^2}{r^{4}}-\frac{r^{2}-r_{g}r}{r^{4}}(r^{2}+L_{z}^{2}-2a\mathcal{E}
L_{z}).
\end{equation}
At the turning points $\dot{r}=0$, the equation $(\ref{4})$ is
quadratic in $\mathcal{E}$ whose solution is
\begin{equation}\label{17}
\mathcal{E}=\frac{aL_{z}r_{g}\pm\sqrt{r^{5}(r-r_{g})+L_{z}^{2}(r^{4}-r^{3}r_{g}+a^{2}r_{g}^{2})}}{r^{3}},
\end{equation}
which gives $\mathcal{E}=V_\text{eff}$, the effective potential. The
condition $\dot{r}=0$ is termed as the turning point because it
gives the location at which an incoming particle turns around from
the neighborhood of the gravitating source \cite{azad}. As we are
considering only the positive energy therefore we will consider only
the positive sign before the square root in equation $(\ref{17})$
for all the further calculation. Equations $(\ref{4})$ and
$(\ref{17})$ hold for equatorial plane only. It can be seen from
$(\ref{17})$ that $\mathcal{E}\rightarrow1$ for
$r\rightarrow\infty$. Therefore the minimum energy for the particle
to escape from the vicinity of black hole is $1$.

Consider a particle in ISCO, where $r_{o}$ is the local minima
(which is also the convolution point) of the effective potential
\cite{14}. The corresponding energy and azimuthal angular momentum
are given by \cite{14,Hobson} after neglecting terms which involving
$a^{2}$ we have
\begin{equation}\label{z1}
L_{zo}=\pm\frac{\sqrt{r_{g}}\bigg(r_{o}\pm a\sqrt{\frac{2r_{g}}{r_{o}}}\bigg)}
       {\sqrt{2r_{o}-3r_{g}\mp 2a\sqrt{\frac{2r_{g}}{r_{o}}}}},
\end{equation}
\begin{equation}\label{10b}
  \mathcal{E}_{o}=\frac{1-\frac{r_{g}}{r}\mp\frac{a}{r}\sqrt{\frac{r_{g}}{2r}}}{\sqrt{1-\frac{3r_{g}}{2r}}\mp\frac{a}{r}\sqrt{\frac{2r_{g}}{r}}}.
\end{equation}
Now consider the particle  in the ISCO which collides with another
incoming particle. After collision between these particles, three
cases are possible for the motion of the particle: (i) bound motion
(ii) capture by the black hole (iii) escape to infinity. The result
will depend on the collision process. For small change in energy and
momentum, orbit of the particle will be slightly perturbed. While
for large change in energy and angular momentum, the particle can
either be captured by black hole or escape to infinity.

After the collision particle should have new values of energy and
momentum $\mathcal{E}$, $L_{z}$ and the total angular momentum
$L^{2}$. We simplify the problem by applying the following
conditions $(i)$ the azimuthal angular momentum is fixed $(ii)$
initial radial velocity remains same after the collision. Under
these conditions only energy of the particle can determine its
motion. After collision particle acquires an escape velocity
$(v_{\bot})$ in orthogonal direction of the equatorial plane
\cite{rong}. The square of total angular momentum of the particle
after collision is given by
\begin{equation}\label{18}
  L^{2}=r^{4}\dot{\theta^{2}}+r^{4}\sin^{2}\theta\dot{\phi}^{2}.
\end{equation}
Putting the value of $\dot{\phi}$ from equation $(\ref{3})$ in
equation $(\ref{18})$ we have
\begin{equation}\label{18a}
  L^{2}=r^{2}v_{\perp}^{2}+\sin^{2}\theta\bigg(\frac{ar_{g}\mathcal{E}_{o}}{r-r_{g}}+\frac{L_{zo}}{\sin^{2}\theta}\bigg)^{2}.
\end{equation}
Here we denote $v\equiv -r\dot{\theta}_{o}$. Note that $L^2$ is not
the integral of motion. It is conserved for $a=0$ i.e. in the
spherically symmetric case. However now the metric is axially
symmetric, therefore only $L_z$ component is conserved. In a flat
spacetime, all three components $L_x$, $L_y$, $L_z$ are conserved,
and so is the square of the total angular momentum. The angular
momentum $L_{zo}$ and energy $\mathcal{E}_{o}$ appearing in
(\ref{18a}) are given by $(\ref{z1})$ and $(\ref{10b})$ which
provide the necessary corrections due to spin of the black hole.

From equations $(\ref{17})$ and $(\ref{18a})$, the angular momentum
and the energy of the particle after the collision becomes
\begin{align}
\label{33}
  L^{2}&=r_{o}^{2}v_{\perp}^{2}+\bigg(\frac{ar_{g}\mathcal{E}_{o}}{r_{o}-r_{g}}+L_{zo}\bigg)^{2},\\
\mathcal{E_\text{new}}&=\frac{aLr_{g}+\sqrt{r_{o}^{5}(r_{o}-r_{g})
  +L^{2}(r_{o}^{4}-r_{o}^{3}r_{g}+a^{2}r_{g}^{2})}}{r_{o}^{3}}.
\end{align}
These values of angular momentum and energy are greater than their
values before the collision. Physically
it means that the energy of the particle exceeds its rest mass
energy. We have mentioned above all the orbits with $\mathcal{E}_\text{new}\geq1$
are unbounded in the sense that particle escapes to infinity.
Conversely for $\mathcal{E}_\text{new}<1$, particle cannot escape to
infinity (the orbits are always bounded).

Therefore particle escapes to infinity if
$\mathcal{E}_\text{new}\geq1$, or
\begin{eqnarray}\label{20b}
  v_{\perp}\geq\pm\frac{r(r_{g}-r)(L_{z}(r-r_{g})+ar_{g}(\mathcal{E}_{o}-1))+\sqrt{r^{2}r_{g}(r-r_{g})^{2}(r^{3}
  +r_{g}(a^{2}-r^{2}-2a^{2}\mathcal{E}_{o}))}}{r^{2}(r-r_{g})^{2}}.
\end{eqnarray}
Particle escape condition is $|v|\geq v_{\perp}$ i.e. the magnitude
of velocity should be greater than any orthogonal velocity.

\section{Charged Particle Around the Slowly Rotating Magnetized Kerr Black Hole}

Here we investigate the motion of a charged particle  (electric
charge $q$) in the presence of  magnetic field in the exterior of
the slowly rotating Kerr black hole. The Killing equation is
\begin{equation}\label{34}
\square \xi^{\mu}=0,
\end{equation}
where $\xi^{\mu}$ is a Killing vector. Note that (\ref{34}) follows
from the result: a Killing vector in a vacuum spacetime generates a
solution of Maxwell equations i.e. $F_{\mu\nu}=-2\xi_{\mu;\nu}$.
From $F^{\mu\nu}_{~~;\nu}=0$, it follows that
$-2\xi^{\mu;\nu}_{~~;\nu}=0$. Thus  (\ref{34}) coincides with the
Maxwell equation for 4-potential $A^{\mu}$ in the Lorentz gauge
$A^{\mu}_{\ \ ;\mu}=0$. The special choice for $A^{\mu}$ is
\cite{6,black}.
\begin{equation}
  A^{\mu}= \Big( a\mathcal{B},0,0,\frac{\mathcal{B}}{2} \Big),
\end{equation}
where $\mathcal{B}$ is the magnetic field strength. The
$4$-potential is invariant under the symmetries which correspond to
the Killing vectors, i.e.,
\begin{equation}
  L_{\xi}A_{\mu}=A_{\mu,\nu}\xi^{\nu}+A_{\nu}\xi^{\nu}_{,\mu}=0.
\end{equation}
A magnetic field vector is defined as
\begin{equation}\label{5}
  \mathcal{B}^{\mu}=-\frac{1}{2}e^{\mu\nu\lambda\sigma}F_{\lambda\sigma}u_{\nu},
\end{equation}
where
\begin{equation}\label{6}
  e^{\mu\nu\lambda\sigma}=\frac{\epsilon^{\mu\nu\lambda\sigma}}{\sqrt{-g}},\ \
  \epsilon_{0123}=1,\ \ g=det(g_{\mu\nu}).
\end{equation}\\
In $(\ref{6})$ $\epsilon^{\mu\nu\lambda\sigma}$ is the Levi Civita
symbol and the Maxwell tensor is defined as
\begin{equation}\label{7}
  F_{\mu\nu}=A_{\nu;\mu}-A_{\mu;\nu}.
\end{equation}
For a local observer at rest we have
\begin{equation}\label{8}
  u^{\mu}=\Big(\frac{1}{\sqrt{(1-\frac{r_{g}}{r})+\frac{4aM\sqrt{(1-\frac{r_{g}}{r})}}{r^{2}\sin\theta}}}
,0,0,\frac{1}{r\sin\theta\sqrt{(1+\frac{4aM}{r^{2}\sin\theta\sqrt{(1-\frac{r_{g}}{r})}})}}\Big).
\end{equation}
From $(\ref{5})-(\ref{8})$ we can obtain the components of magnetic
field
\begin{eqnarray}\label{19}
  \mathcal{B}^{\mu}&=&\mathcal{B}\Big(0,\cos\theta\Big(\frac{(1-\frac{r_{g}}{r})}{\sqrt{(1-\frac{r_{g}}{r})+
  \frac{2r_{g}a\sqrt{(1-\frac{r_{g}}{r})}}{r^{2}\sin\theta}}}\Big)+\frac{r_{g}a\sin\theta\cos\theta}{r}
  \Big(\frac{(1-\frac{r_{g}}{r})}{r\sin\theta\sqrt{(1+\frac{2r_{g}a}{r^{2}\sin\theta\sqrt{(1-\frac{r_{g}}{r})}})}}
   \Big),
  \nonumber\\&&
  ,-\frac{\sin\theta(1-\frac{r_{g}}{r})}{r\sqrt{(1-\frac{r_{g}}{r})+\frac{2r_{g}a\sqrt{(1-\frac{r_{g}}{r})}}
  {r^{2}\sin\theta}}},0\Big).
\end{eqnarray}
For the equatorial plane only, the third component of the magnetic
field  will survive. Hence equation $(\ref{19})$ becomes
\begin{eqnarray}\label{19a}
  \mathcal{B}^{\mu}&=&\mathcal{B}\Big(0,0,
  -\frac{(1-\frac{r_{g}}{r})}{r\sqrt{(1-\frac{r_{g}}{r})+\frac{2r_{g}a\sqrt{(1-\frac{r_{g}}{r})}}
  {r^{2}}}},0\Big).
\end{eqnarray}

The Lagrangian of the particle of mass $m$ and charge $q$ moving in
an external magnetic field in a curved spacetime is \cite{12}
\begin{equation}\label{10}
  \mathcal{L}=\frac{1}{2}g_{\mu\nu}\dot{x}^{\mu}\dot{x}^{\nu}+\frac{qA_{\mu}}{m}\dot{x}^{\mu},
\end{equation}
and generalized 4-momentum of the particle is
$P_{\mu}=mu_{\mu}+qA_{\mu}$. The constants of motion are
\begin{align}\label{1.5}
  \dot{t}&=\frac{r^{3}\mathcal{E}+aL_{z}r_{g}}{r^{2}(r-r_{g})}-2aB,
  \nonumber\\
  \dot{\phi}&=\frac{1}{r^{2}}\bigg(\frac{ar_{g}\mathcal{E}}{(r-r_{g})}+\frac{L_{z}}{\sin^{2}\theta}\bigg)-B.
\end{align}
For the equatorial plane $\theta=\frac{\pi}{2}$ the above integrals of motion become
\begin{align}\label{1.5d2}
  \dot{t}&=\frac{r^{3}\mathcal{E}+aL_{z}r_{g}}{r^{2}(r-r_{g})}-2aB,
  \nonumber\\
  \dot{\phi}&=\frac{1}{r^{2}}\big(\frac{ar_{g}\mathcal{E}}{(r-r_{g})}+L_{z}\big)-B.
\end{align}
Here we denote
\begin{equation}\label{12}
  B\equiv\frac{q\mathcal{B}}{2m}.
\end{equation}
After putting the value of $\dot{t}$ and $\dot{\phi}$ and neglecting
the terms involving $a^{2}$, Eq. (\ref{10}) yields
\begin{eqnarray}\label{16a}
  \mathcal{L}&=&\frac{1}{2r^{2}(r-r_{g})^{2}}\bigg[4Br^{2}L_{z}(r_{g}-r)+L_{z}^{2}(r_{g}-r)+
  \nonumber\\&&
  r^{2}\big(Br_{g}(3Br^{2}+2a\mathcal{E})r(\mathcal{E}^{2}-3B^{2}r^{2}-\dot{r}^{2})\big)\bigg].
\end{eqnarray}
By using the above Lagrangian in Euler-Lagrange equation which is
defined as
\begin{equation}\label{16b}
  \frac{d}{d\tau}\big(\frac{\partial\mathcal{L}}{\partial\dot{x}}\big)-\frac{\partial\mathcal{L}}{\partial{x}}=0,
\end{equation}
we get
\begin{eqnarray}\label{14}
  \ddot{r}&=&\frac{Ba\mathcal{E}r_{g}}{r(r-r_{g})}
  +\frac{1}{2r^{4}(r-r_{g})}\bigg[6B^{2}r^{6}-2L_{z}^{2}(r-r_{g})^{2}
  \nonumber\\&&
  +r^{3}r_{g}(-\mathcal{E}^{2}+6B^{2}rr_{g}+\dot{r}^{2}-12B^{2}r^{2})\bigg].
\end{eqnarray}
Following the procedure of section II, using the normalization
condition, $u^{\mu}u_{\mu}=1$ and putting the value of new constants
of motion $(\ref{1.5d2})$,
 we obtain
\begin{eqnarray}\label{15}
  \mathcal{E}&=&\frac{1}{r_{o}^{6}(r_{o}-r_{g})}\bigg[2aBr_{o}^{7}+ar_{g}r_{o}^{3}\big(2Br_{o}^{2}(r_{g}-2r_{o})+L_{z}(r_{g}-r_{o})\big)
  \nonumber\\&&
  \pm\bigg(a^{2}r_{o}^{6}(r_{o}-r_{g})^{2}\big(r_{g}(L_{z}+2Br_{o}^{2})-2Br_{o}^{3}\big)^{2}+
  \nonumber\\&&
  r_{o}^{9}(r_{o}-r_{g})^{3}\big(r_{o}^{2}+(L_{z}+Br_{o}^{2})^{2}\big)\bigg)^{\frac{1}{2}}
  \bigg].
\end{eqnarray}
If $(\ref{15})$ is satisfied initially (at the time of collision),
 then it is always valid (throughout the motion), provided that
 $r(\tau)$ is controlled by $(\ref{14})$.

The system $(\ref{10})-(\ref{15})$ is invariant with respect to
reflection $(\theta\rightarrow\pi-\theta)$. This transformation
retains the initial position of the particle and changes
$(v_{\perp}\rightarrow-v_{\perp})$ as it is defined,
$(v_{\perp}\equiv-r\dot{\theta_{o}})$. Therefore, it is sufficient
to consider only the positive value of $(v_{\perp})$.

\section{Dimensionless Form of the Dynamical Equations}

To perform the numerical analysis, it is convenient to convert
equations (\ref{14}) and (\ref{15}) to dimensionless form by
introducing the following dimensionless quantities
\begin{equation}
  \sigma=\frac{\tau}{r_{g}},\ \rho=\frac{r}{r_{g}},\ \ell=\frac{L_{z}}{r_{g}},\ b=Br_{g}.
\end{equation}

Equation $(\ref{15})$ now becomes
\begin{eqnarray}\label{21}
\mathcal{E}_{o}&=&\frac{1}{\rho_{o}^{6}(\rho_{o}-1)}\bigg[a\rho_{o}^{3}(1-\rho_{o})(\ell-2b\rho_{o}^{2}(\rho_{o}-1))
         \nonumber\\&&
         +\bigg(\rho_{o}^{6}(\rho_{o}-1)^{2}(a^{2}\big(\ell-2b\rho_{o}^{2}(\rho_{o}-1)\big)^{2})
         \nonumber\\&&
         +\rho_{o}^{3}(\rho_{o}-1)\big(\rho_{o}^{2}+(\ell+b\rho_{o}^{2})^{2}\big)
         \bigg)^{\frac{1}{2}}\bigg].
\end{eqnarray}
The magnetic field is zero at $\rho\rightarrow\infty$. Therefore from the equation $(\ref{21})$ as  $\rho\rightarrow\infty$ then $\mathcal{E}\rightarrow1$.

\begin{figure}[!ht]
\centering
\includegraphics[width=10cm]{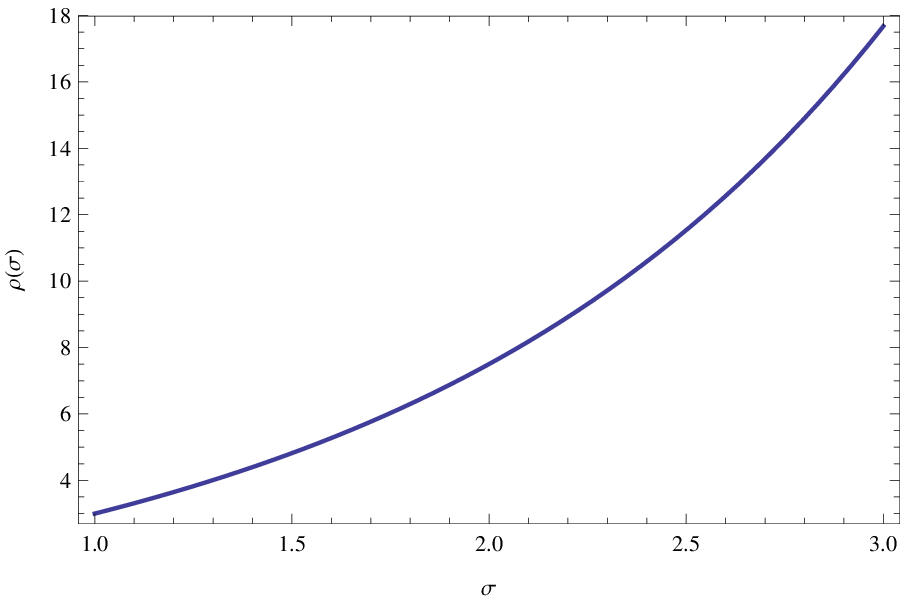}
\caption{Figure shows the graph for $\rho(\sigma)$ vs $\sigma$. Here
$\mathcal{E}=1, q=1, b=0.5, \ell=2,$ and $a=0.1$.}\label{f31}
\end{figure}

\begin{figure}[!ht]
\centering
\includegraphics[width=10cm]{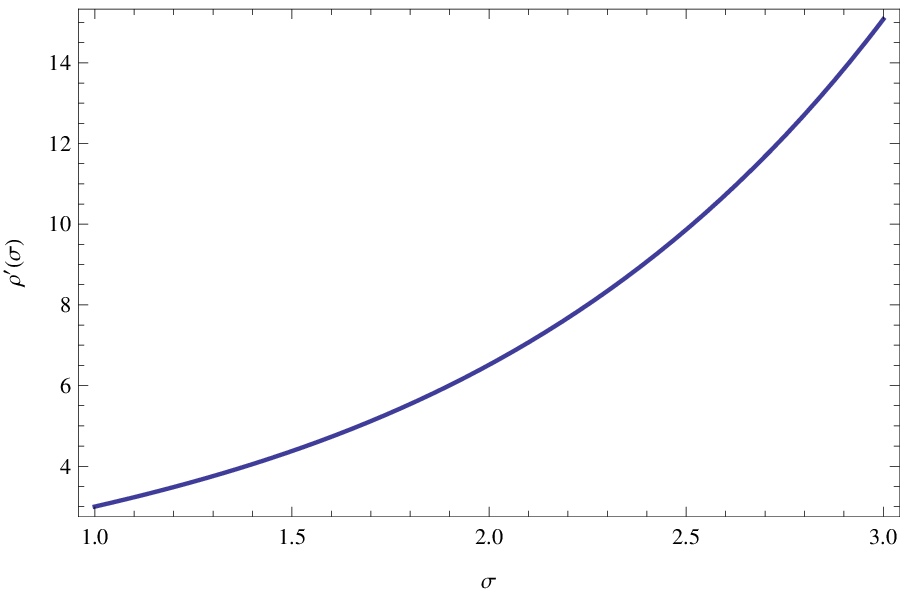}
\caption{Figure shows the graph for $\rho'(\sigma)$ (radial
velocity) vs $\sigma$. Here $\mathcal{E}=1, q=1, b=0.5, \ell=2,$ and
$a=0.1$.}\label{f32}
\end{figure}

Dimensionless form of equation $(\ref{14})$ is
\begin{eqnarray}\label{1.4}
  \frac{d^{2}\rho}{d\sigma^{2}}&=&\frac{1}{2\rho^{4}(\rho-1)}\big[\rho^{3}\big(2a\mathcal{E}b+6\mathcal{E}^{2}b^{2}\rho(\rho-1)^{2}\big)
  -2\ell(\rho-1)^{2}+\rho^{3}\frac{d\rho}{d\sigma}\big].
\end{eqnarray}
We solved the equation $(\ref{1.4})$ numerically by using the built
in command NDSolve  of  Mathematica. As ISCO exists at $r=3r_{g}$,
and using $\rho=\frac{r}{r_{g}}$ and $\sigma=\frac{\tau}{r_{g}}$,
our initial conditions for solving $(\ref{1.4})$ become $\rho(1)=3$
and $\dot{\rho}(1)=3$. We get the interpolating function
$\rho(\sigma)$ as the solution of the equation $(\ref{1.4})$ which
we plotted in figure \ref{f31} against $\sigma$. In figure \ref{f32}
we have plotted the radial velocity (derivative of the interpolating
function) vs $\sigma$ which shows that the particle will escape to
infinity according to the initial conditions.

As is the case of a neutral particle, we assume that the collision
does not change the azimuthal angular momentum of the particle but
it changes the transverse velocity $v>0$. Due to this, the angular
momentum and the energy of the particle will change as
$\ell\rightarrow\ell_{t}$ and
$\mathcal{E}_{o}\rightarrow\mathcal{E}$ respectively which is given
by
\begin{equation}\label{20a}
  \ell_{t}^{2}=\rho^{2}v_{\perp}^{2}+\rho^{4}\bigg[\frac{1}{\rho^{2}}\bigg(\frac{a\mathcal{E}_{o}}{2(\rho-1)}+\ell\bigg)-b\bigg]^{2},
\end{equation}
\begin{eqnarray}\label{20}
\mathcal{E}&=&\frac{1}{\rho_{o}^{6}(\rho_{o}-1)}\bigg[a\rho_{o}^{3}(1-\rho_{o})(\ell_{t}-2b\rho_{o}^{2}(\rho_{o}-1))
         \nonumber\\&&
         +\bigg(\rho_{o}^{6}(\rho_{o}-1)^{2}(a^{2}\big(\ell_{t}-2b\rho_{o}^{2}(\rho_{o}-1)\big)^{2})
         \nonumber\\&&
         +\rho_{o}^{3}(\rho_{o}-1)\big(\rho_{o}^{2}+(\ell_{t}+b\rho_{o}^{2})^{2}\big)
         \bigg)^{\frac{1}{2}}\bigg].
\end{eqnarray}
Here $\ell_{t}$ is the dimensionless form of $L$ given by equation $(\ref{18a})$.
For the unbound motion
$\mathcal{E}\geq1$.  By solving $(\ref{20})$ and putting
$\mathcal{E}=1$, we get escape velocity of the
particle as given below
\begin{eqnarray}\label{31}
  v_{\perp}&=&\pm \frac{1}{4\rho^{2}(\rho-1)}\bigg[4(\rho-1)\big[\sqrt{a^{2}\rho^{2}+\rho^{4}(\rho-1)-2ab\rho^{4}(\rho-1)(2\rho-1)}
  \nonumber\\&&
  -a\rho+\rho(\rho-1)(b\rho^{2}+(\ell-b\rho^{2})^{2})\big]+a\rho\mathcal{E}_{o}\big[4(\rho-1)(\ell-b\rho^{2})\big]
  \bigg]
\end{eqnarray}

We now discuss the behavior of the particle when it escapes to
asymptotic infinity. For simplicity we consider the particle
initially in ISCO. The parameter $\ell$ and $b$ are defined in term
of $\rho_{o}$ and only $\mathcal{E}$ specifies the motion of the
particle. We can express the parameters $\ell$ and $b$ in term of
$\rho_{o}$ by simultaneously solving  the equations
$\frac{d\mathcal{E}_{o}}{d\rho}=0$, and
$\frac{d^{2}\mathcal{E}_{o}}{d\rho^{2}}=0$, for $\ell$ and $b$.
 But the first derivative  and second derivative  of effective potential are
very complicated and we cannot find the explicit expression for $\ell$ and $b$ in term of $\rho$.

\section{Trajectories for Escape Energy}

\begin{figure}[!ht]
\centering
\includegraphics[width=11cm]{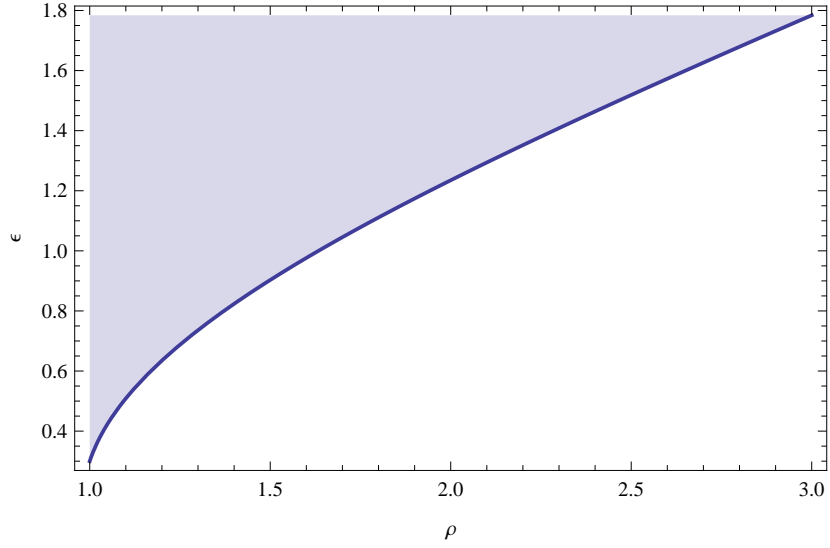}
\caption{In this figure we plot Effective potential $\mathcal{E}$ as
a function of $\rho$ for $\ell=5$, $b=0.5$ and $a=0.1$. }\label{f15}
\end{figure}

\begin{figure}[!ht]
\centering
\includegraphics[width=11cm]{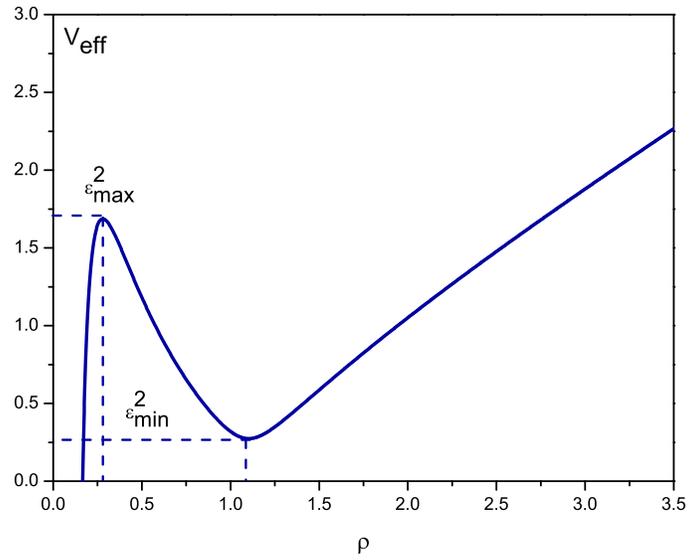}
\caption{
Here we plot the effective potential against $\rho$  for $\ell=20$,
$b=0.5$, and $a=0.1$. In this figure $\mathcal{E}_{max}$ corresponds
to unstable circular orbit and $\mathcal{E}_{min}$ corresponds to
stable circular orbit.} \label{f28}
\end{figure}

\begin{figure}[t]
\centering
\includegraphics[width=11cm]{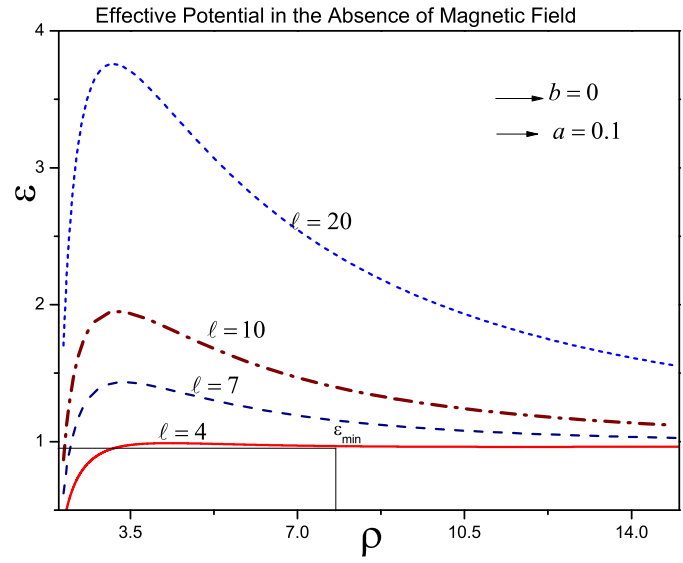}
\vspace{-.8cm}\caption{The effective potential $\mathcal{E}$ is plotted as a function of radial coordinate $\rho$ for different values of angular momentum $\ell$.}\label{p6}
\end{figure}

\begin{figure}[t]
\centering
\includegraphics[width=11cm]{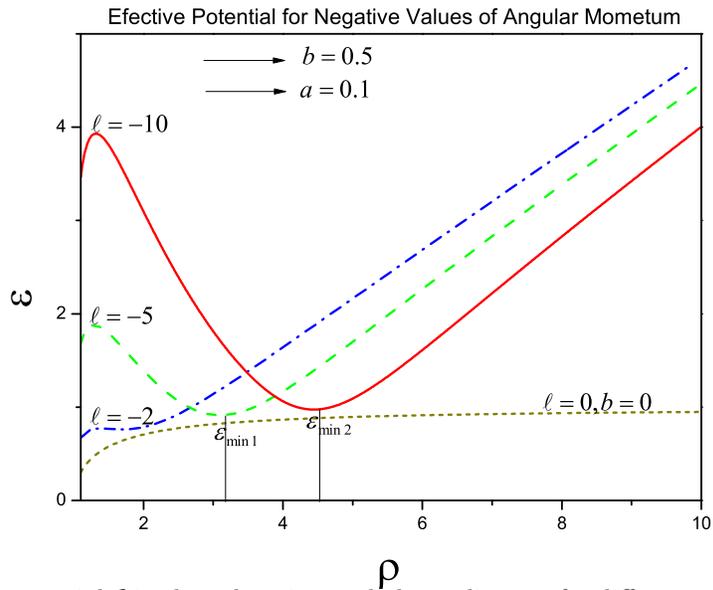}
\vspace{-.8cm}\caption{The effective potential $\mathcal{E}$ is plotted against radial coordinate $\rho$ for different values of negative angular momentum $\ell$.}\label{PLD}
\end{figure}

\begin{figure}[t]
\centering
\includegraphics[width=11cm]{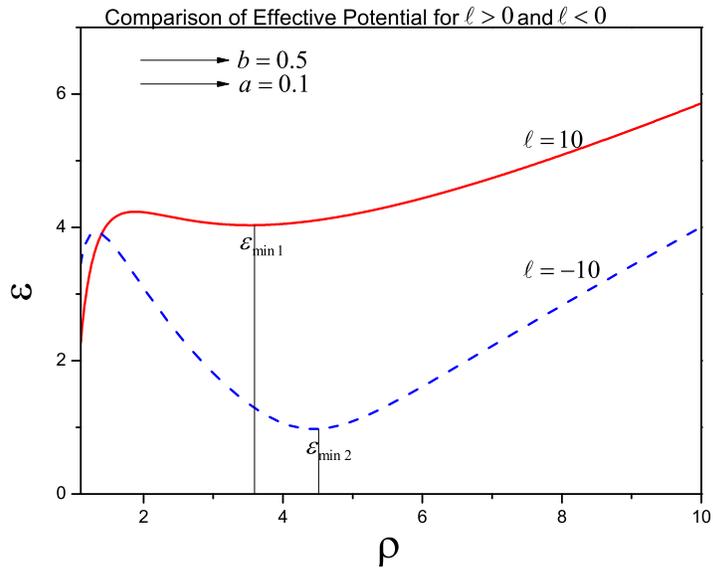}
\vspace{-.8cm}\caption{In this figure we have plotted the effective potential $\mathcal{E}$  vs $\rho$ for $\ell=-10$ and $\ell=10$.}\label{CP}
\end{figure}

\begin{figure}[h]
\centering
\includegraphics[width=11cm]{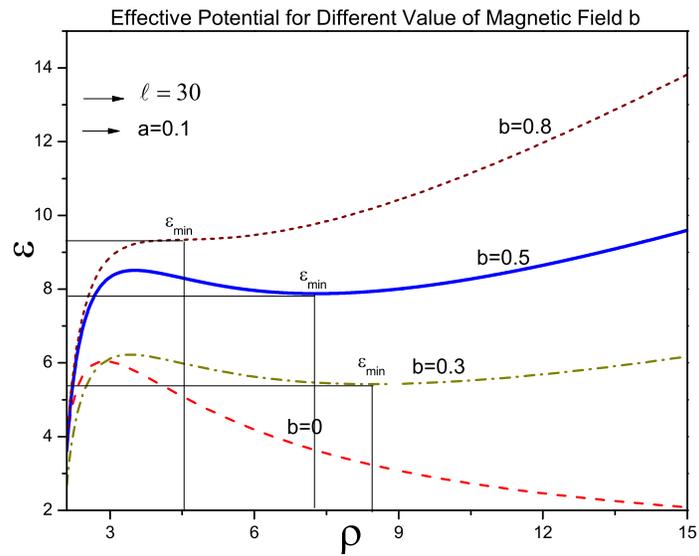}
\vspace{-.8cm}\caption{The effective potential $\mathcal{E}$ against $\rho$ for different values of magnetic field.}\label{p3}
\end{figure}

Here we investigate the dynamics of particle for the positive energy
 $\mathcal{E}_{+}$. Particles with negative energy exist only
inside the static limit surface $(r_{st}=2m)$ orbiting in the
retrograde orbits and do not have the chance to escape.  The
equation for the rotational (angular) variable $\phi$ is
\begin{equation}\label{25}
\frac{d\phi}{d\sigma}=\frac{\ell}{\rho^{2}}-b+\frac{a\mathcal{E}}{\rho^{3}(1-\rho)}.
\end{equation}
The Lorentz force acting on the massive charged particle is
attractive when $d\phi/d\sigma<0$ and vice versa. All the figures
(\ref{f15}-\ref{p3}) correspond to Eq. $(\ref{21})$. In
figure-\ref{f15}, the shaded region corresponds to unbound motion
while the unshaded region refers to bounded trajectories of the
particle. The curved line represents the minimum energy required for
the particle to escape form the vicinity of the black hole. It can
be seen from figure-\ref{f28} that for large values of angular
momentum, the  plot is similar to  the effective potential of
Schwarzschild  black hole \cite{11}.  In figure $\ref{f28}$,
$\mathcal{E}_\text{max}$ corresponds to unstable circular orbit and
$\mathcal{E}_\text{min}$ refers to ISCO.

The effective potential $\mathcal{E}$ of a particle moving in a
slowly rotating Kerr spacetime is plotted as a function of radial
coordinate $\rho$ for different values of angular momentum $\ell$ in
figure $\ref{p6}$. We can see from figure $\ref{p6}$ that for large
value of angular momentum, the maxima is shifting upward. For a
particle to be captured by the black hole it is required that the
energy which should be greater then this maxima. If its energy is
less than this maxima there are two possibilities for a particle
either it will escape to infinity or it might start moving in ISCO.
If energy of the particle $\mathcal{E}<1$ then it will stay in some
stable orbit and if $\mathcal{E}>1$ then it will escape to infinity.
In figure \ref{p6} we plotted effective potential against $\rho$ for
different value of angular momentum. For $\ell>0$ the Lorentz force
is repulsive. Hence it can be concluded from  figure \ref{p6} that
the possibility of a particle to escape after collision from the
vicinity of the black hole is greater for larger value of $\ell_{+}$
as compare to the lesser value of it. For $\ell<0$ the lorentz force
is attractive. Therefore, the possibility of a particle to escape
after collision is less for larger value of $\ell_{-}$ as compared
to smaller value of $\ell_{-}$, represented in figure \ref{PLD}. The
graph for $\ell=0$ and $b=0$ in figure \ref{PLD} corresponds to
photon as there is no stable region. Moreover, we compare the
effective potential for $\ell=10$ and $\ell=-10$ in figure \ref{CP}.
It can be seen that the stability is larger for $\ell=-10$.
Therefore it is concluded that for the attractive Lorentz force
$(\ell=-10)$, particle required more energy to escape. It can be
seen from  figure $\ref{p3}$  that with the increase in the strength
of magnetic field, the local minima of the effective potential is
shifting toward the horizon. This local minima corresponds to ISCO,
which is in agreement with the result of \cite{6}.

\section{Trajectories for Escape Velocity}

\begin{figure}[!ht]
\centering
\includegraphics[width=9cm]{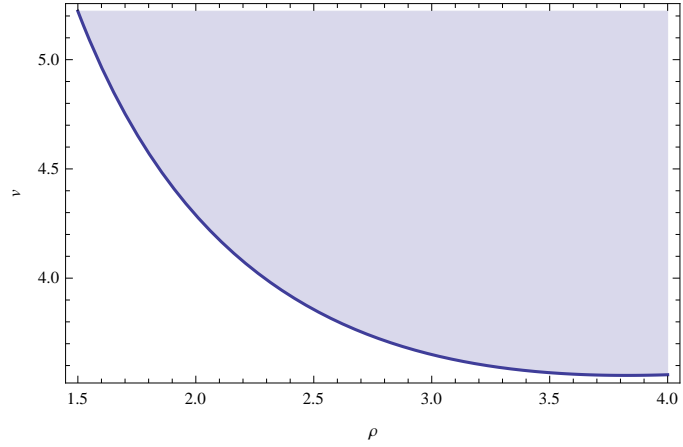}
\caption{Here we have plotted the escape velocity against $\rho$
for $\ell=5$, $b=0.5$ and $a=0.1$.}\label{f16}
\end{figure}

\begin{figure}[!ht]
\centering
\includegraphics[width=9cm]{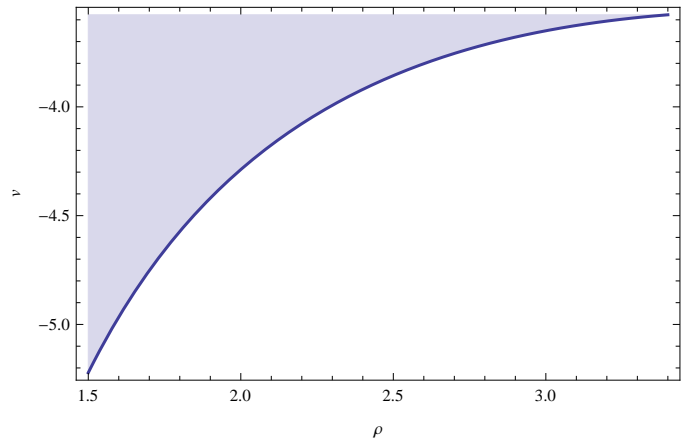}
\caption{In this figure we have plotted the escape velocity against $\rho$
for $\ell=5$, $b=0.5$ and $a=0.1$.}\label{f17}
\end{figure}

\begin{figure}[!ht]
\centering
\includegraphics[width=10cm]{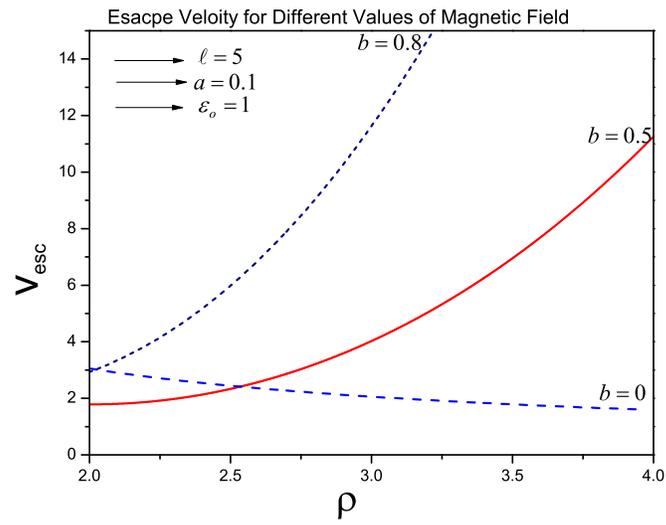}
\caption{Escape velocity $v_{esc}$ against $\rho$ for different values of magnetic field $b$.}\label{v1}
\end{figure}

\begin{figure}[!ht]
\centering
\includegraphics[width=10cm]{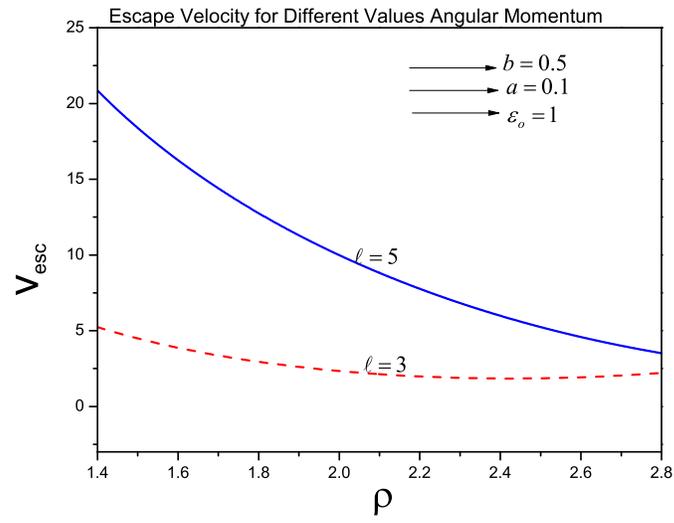}
\caption{Escape velocity $v_{esc}$ against $\rho$ for different values of angular momentum $\ell$.}\label{v2}
\end{figure}

For all the figures of escape velocity we have denoted
$v_{\perp}\equiv v_\text{esc}$. From Eq. $(\ref{20})$ we calculate
the escape velocity by substituting $\mathcal{E}=1$. Figures
\ref{f16}-\ref{v2} correspond to Eq. $(\ref{31})$. In figure
\ref{f16}, the shaded region corresponds to escape velocity of the
particle and the solid curve represents the minimum velocity
required to escape from the vicinity of the black hole to infinity.
The unshaded region represents the bound motion around the black
hole. In figure \ref{f17} the shaded region corresponds to escape
velocity of the particle  and the solid curve represents the minimum
velocity required to escape from the vicinity of the black hole. The
unshaded region represents the bound motion around the black hole.

In figure $\ref{v1}$ we plotted escape velocity of a particle moving
in ISCO as a function of radial coordinate $\rho$ for different
values of magnetic field $b$. It can be seen from figure $\ref{v1}$
that due to the presence of magnetic field in the vicinity of black
hole escape velocity of the particle increases. Therefore we can say
that in the presence of magnetic field $b$, the possibilities of the
particle to escape is greater then the case when magnetic field is
absent i.e. $b=0$. We plotted the escape velocity against $\rho$ in
figure $\ref{v2}$ for different values of angular momentum $\ell$.
We can see from the figure $\ref{v2}$ that the the escape velocity
is increasing for large value of $\ell$. Hence we can conclude that
if particle has larger value of angular momentum $\ell$ then it can
easily escape to infinity as compared to the particle with smaller
value of angular momentum $\ell$ regardless of the magnetic field.

\section{Discussion}

We have studied the dynamics of  a neutral and a charged particle
around the slowly rotating Kerr black hole which is immersed in a
magnetic field. Therefore the particle is under the influence of
both gravitational and electromagnetic forces.  We have obtained
equations of motion by using Lagrangian formalism. We have derived
the expression for magnetic field present in the vicinity of slowly
rotating Kerr black hole.  We have calculated the minimum energy for
a particle to escape from ISCO to infinity. With zero spin i.e.
$a=0$, our results reduce to the case of the Schwarzschild black
hole \cite{13}.

The behavior of effective potential and escape velocity against
magnetic field and angular momentum are discussed in detail. It is
shown in figures \ref{f28}, \ref{f16} and \ref{f17} under what
conditions particle can escape from the vicinity of the  black hole
to spatial infinity. For larger values of the angular momentum,
behavior of the effective potential is similar to that of the
Schwarzschild black hole \cite{13}. It is concluded that magnetic
field largely effects the motion of the particle in the vicinity of
the black hole. This effect decreases far away from the black hole.
It is found that as the value of magnetic field parameter is
increased, the local minima of effective potential shifted towards
the horizon, as shown in figure $\ref{p3}$. This indicates that the
ISCO shrinks as strength of magnetic field increases. It is
concluded from the figures $\ref{p6}$ and $\ref{v2}$ that if
particle has large value angular momentum $\ell_{+}$ then it can
escape easily as compare to particle with smaller angular momentum
$\ell_{+}$. Figure \ref{CP} shows that for attractive Lorentz force
$(\ell_{-})$ the stability is larger in comparison with repulsive
Lorentz force $(\ell_{+})$.

Escape velocity $v_{esc}$, for different values of magnetic field
$b$ is plotted in figure \ref{v1}. It is found that due to the
presence of magnetic field in the vicinity of black hole escape
velocity of the particle increases. Therefore we found that the
possibility of the particle to escape from the vicinity of black
hole to infinity is greater in the presence of magnetic field  as
compared to the case when magnetic field is absent $b=0$.

\subsection*{Acknowledgment}
M. Jamil and S. Hussain would like to thank the Higher Education
Commission, Islamabad, Pakistan for providing financial support
under project grant no. 20-2166.

\pagebreak




\end{document}